# Action spectroscopy of gas-phase carboxylate anions by multiple photon IR electron detachment/attachment

Jeffrey D. Steill and Jos Oomens*

FOM Institute for Plasma Physics 'Rijnhuizen', Edisonbaan 14, 3439MN Nieuwegein,

The Netherlands

*Corresponding author: joso@rijnhuizen.nl

**Abstract.**

We report on a form of gas-phase anion action spectroscopy based on infrared multiple photon electron detachment and subsequent capture of the free electrons by a neutral electron scavenger in a Fourier Transform Ion Cyclotron Resonance (FTICR) mass spectrometer. This method allows one to obtain background-free spectra of strongly bound anions, for which no dissociation channels are observed. The first gas-phase spectra of acetate and propionate are presented using $SF_6$ as electron scavenger and a free electron laser as source of intense and tunable infrared radiation. To validate the method, we compare infrared spectra obtained through multiple photon electron detachment/attachment and multiple photon dissociation for the benzoate anion. In addition, different electron acceptors are used, comparing both associative and dissociative electron capture. The relative energies of dissociation (by $CO_2$ loss) and electron detachment are investigated for all three anions by DFT and CCSD(T) methods. DFT calculations are also employed to predict vibrational frequencies, which provide a good fit to the infrared spectra observed. The frequencies of the symmetric and antisymmetric carboxylate stretching modes for the aliphatic carboxylates are compared to those previously observed in condensed-phase IR spectra and to those reported for gas-phase benzoate, showing a strong influence of the solution environment and a slight substituent effect on the antisymmetric stretch.

**I. Introduction.**

Infrared spectroscopy of gas-phase molecular ions through action spectroscopy techniques has seen rapid growth over the past decades and two of the most prominent techniques are vibrational predissociation spectroscopy using the "messenger-atom" method and infrared multiple photon dissociation (IRMPD) spectroscopy in ion traps and guides (see e.g. several recent review articles).[1-4] A particular distinction between these two experimental methods lies in the jet-cooled environment of vibrational predissociation spectroscopy in contrast to IRMPD spectroscopy being mostly performed at thermal energies, which makes this latter method particularly suitable for use with electrospray ionization (ESI). IRMPD methods require high intensity infrared lasers and rapid intramolecular redistribution of vibrational energy (IVR) to allow for multiple-photon absorption at IR-active resonances, resulting in internal energies sufficiently large to break covalent bonds. For anions, the increased internal energy can result in either dissociation or electron detachment,[5-7] since detachment energies are typically comparable to bond dissociation energies. For anions that lack an easily energetically or entropically accessible dissociation channel, the electron detachment channel may dominate. For instance, IRMPD spectroscopy of small, tightly bound molecules lacking the many vibrational degrees of freedom required for efficient IVR is often not possible. Thus, small molecular anions are an interesting class of gas-phase ions to be studied with IR multiple photon detachment spectroscopy.

Several infrared action spectroscopy methods based on electron ejection have been developed. Infrared autodetachment spectroscopy has been applied to study the spectroscopy of anions with relatively low electron affinities. Excitation of a vibrational band, usually in the hydrogen stretching region, which couples to the electron continuum leads to resonances in the autodetachment of the electron. Systems studied using this method include the imidogen anion[8] (NH$^-$), the nitromethane anion[9] and anionic water clusters.[10] IR multiple photon induced ejection of electrons has been observed for neutral species with a dissociation threshold higher than the ionization potential, such as fullerenes and metal clusters.[11] Infrared multiple photon detachment of molecular anions has also been applied.[5,12] Brauman and coworkers reported infrared multiple photon electron detachment of the benzyl anion,[5] where the detached electron was detected using $CCl_4$ as electron scavenger forming $Cl^-$. The method could however not be used as a spectroscopic technique since no sharp spectral features were observed. Also the experiment by Beauchamp[12] on isomers of $C_7H_7^-$ appears to suffer from the limited tuning range of a $CO_2$ laser, which results effectively in only one observable IR band for two isomers. Recently, we reported the IRMPD spectrum of $SF_5^-$ obtained in a Fourier Transform ion-cyclotron resonance (FTICR) mass spectrometer.[13] Upon resonance, formation of $SF_6^-$ was observed due to capture of the detached electrons by residual neutral $SF_6$ in the ICR cell. Detection of photodetachment in FTICR mass spectrometers is facilitated by the use of electron attaching molecules, since the cyclotron frequency of free electrons is too high for conventional detection. The use of $SF_6$, $CCl_4$ and other molecules with large low-energy free electron capture cross-sections is often referred to as the electron scavenger technique and a thorough description of this method is given by Kohan *et al*.[14] Application of this method to the infrared is limited, however. It is anticipated that infrared irradiation of molecules of small to moderate size with comparable dissociation and detachment energies may show a competition between the two channels, and this is explored here as a means of gas-phase infrared action spectroscopy. As an example, we study the infrared spectra of the small aliphatic carboxylate anions acetate and propionate, as well as that of the benzoate anion, which was recently investigated with IRMPD spectroscopy.[15]

The IRMPD spectrum of benzoate was obtained by monitoring the appearance of the phenide anion, $C_6H_5^-$, and DFT calculations provide a good fit to the spectrum. The splitting between the symmetric and antisymmetric carboxylate stretching modes, $\Delta\nu_{a-s}$, is often used as a measure of the relative strength of counterion binding to the carboxylate moiety in gas-phase clusters, solutions and salts. Comparing the benzoate frequencies to those of aliphatic carboxylates can provide some measure of the effect of resonance stabilization on the carboxylate stretching frequencies. However, the small size and hence the low state densities of the aliphatic carboxylate anions acetate and propionate make it impossible to measure their infrared spectra by the same methods as the benzoate anion. Condensed-phase IR spectra of the acetate[16-18] and propionate[16,19] anions as sodium salts and in aqueous solution have been reported but a gas-phase spectrum is required for determination of the inherent, unperturbed vibrational frequencies of these ions.

Gas-phase studies of the acetate, propionate and benzoate anions have been performed using photoelectron spectroscopy. Woo *et al.* determined the adiabatic electron detachment energy of the benzoate anion to be 3.4-3.6 eV.[20] Similarly, studies on the acetate anion[21-24] yield an adiabatic electron affinity of the acetyloxyl radical of 3.47 eV [21] and a low-temperature improved experimental value of 3.25 eV.[24] The propionate anion was shown to have an adiabatic electron detachment energy quite similar to that of the acetate anion.[23]

Electron detachment may compete with dissociation by $CO_2$ loss. Although the dynamics of intramolecular energy transfer likely play a primary role, the thermodynamic values provide a foundation for discussion of the competition between detachment and dissociation. Using thermodynamic cycles based on experimental values, Lu and Continetti[21] determined the dissociation energy of acetate as 2.48 eV, which compares favorably to the collision-induced dissociation value of 2.5 eV.[25] Thus, while the energy of $CO_2$ loss is smaller than the electron detachment energy, the two processes are similar enough in energy that they can be expected to compete upon infrared irradiation.

**II. Experimental Methods.**

Benzoate, propionate and acetate ions are generated by electrospray ionization of 1mM solutions of the corresponding acids in a 80:20 MeOH:$H_2O$ mixture. A small amount of NaOH was added to enhance the deprotonation of the analytes. Ionization with a Micromass "Z-spray" electrospray ionization (ESI) source and subsequent storage and isolation in the Penning trap of a homebuilt FTICR mass spectrometer[26] has been described previously.[27] Infrared radiation in the range of 5.5 to 11 μm is generated by the free-electron laser (FEL) FELIX located at our institute.[28] The ions are irradiated for 3 seconds at a macropulse repetition rate of 5 Hz, with 3 mass spectra averaged per spectral data point. The bandwidth of the FEL is less than 1% of the central wavelength over the range of 1800 to 600 cm$^{-1}$. The background pressure in the cell with addition of the background electron scavenger gases SF$_6$ and CH$_3$I was approximately 2x10$^{-7}$ Torr. Detachment and dissociation yield spectra are produced by dividing the product ion signal by the total ion signal, i.e. the sum of parent and product ions, as a function of FEL wavelength.

Minimum energy geometries and harmonic vibrational frequencies were computed by DFT methods implemented in Gaussion03.[25] All structures were verified to be true minima on the potential energy surface by harmonic frequency calculations. For acetate and propionate the optimization and harmonic vibrational frequency calculations were performed at multiple conformations to verify minimum energy configurations. Energies of electron detachment and dissociation by CO$_2$ loss for all carboxylates were investigated using DFT (B3LYP) and Coupled-Cluster (CCSD(T)) methods. The carboxylate anions were calculated as closed-shell singlet states, and the neutral radical species were calculated as open-shell doublet states using spin-unrestricted methods. For the CCSD(T) calculations, energies are computed at the B3LYP/aug-cc-pVDZ geometries. Zero-point energy corrections are applied to the thermodynamic energy (zero Kelvin) at the level of theory used, except for CCSD(T) values, which are corrected using the unscaled B3LYP/aug-cc-pVDZ harmonic frequencies. Aqueous solution-phase vibrational frequencies were calculated using the Onsager dipole-dipole sphere self-consistent reaction field model implemented with a spherical radius of 3.44, 3.74 and 4.30 Å for acetate, propionate and benzoate, respectively.

**III. Results and Discussion.**

**A. Benzoate Anion Dissociation and Detachment.**

In a previous study[16], the IRMPD spectrum of benzoate ($C_6H_5COO^-$) was found to correlate well with DFT scaled harmonic frequency calculations. Addition of a small amount of $SF_6$ background gas to the ICR cell reveals an additional photodecomposition channel: in addition to dissociation, detachment is observed via the $SF_6^-$ anion due to associative electron attachment (Figure 1). The spectral bands from the photodetachment channel clearly correlate with those observed from depletion of the $C_6H_5COO^-$ intensity and dissociation into the $C_6H_5^-$ channel. The small discrepancy in intensity of the band at 800 cm$^{-1}$ is possibly a manifestation of different dissociation and detachment dynamics, but considering the much larger fraction of dissociation compared to detachment, this may also be simply a factor of a lower signal-to-noise ratio.

To verify that the phenomenon observed is in fact due to electron capture by the background gas, a different electron capture agent, methyl iodide, is used over a limited spectral region (mass spectra are shown in the Supporting Information). Dissociative electron attachment to $CH_3I$ producing $I^-$ is shown overlaid with the associative electron attachment to $SF_6$ in Figure 2. The electron capture cross section of methyl iodide is less than that of $SF_6$ at low electron energies, but little difference is observed in the scavenger yield since the photodetached electrons remain trapped in the ICR cell, resulting in effectively equivalent collection efficiencies. The carboxylate vibrations clearly correlate for both electron scavengers, demonstrating the generality of the method.

Scavengers resulting in an atomic anion from dissociative electron attachment, such as $CH_3I$ and $CCl_4$, have the advantage that the reaction product does not interact with the IR radiation, whereas anionic $SF_6$ could undergo dissociation as a result of IR absorption. We have recently measured the IRMPD spectrum[13] of $SF_6^-$ and the only substantial dissociation was found in the region 600-800 cm$^{-1}$, producing essentially only $SF_5^-$. Experimentally, the use of $SF_6$ as was often found to give more stable signals because of its higher vapor pressure and inherently more constant background pressure over the course of an IR scan. In the present experiments, no appreciable $SF_5^-$ signal was observed, which suggests that the photodetached electrons are of very low energy.

**B. Acetate and Propionate Detachment Spectroscopy.**

Unlike the benzoate anion, irradiation of acetate and propionate anions does not show any detectable dissociation products under our experimental conditions. However, as shown in Figures 3 and 4, appearance of $SF_6^-$ resulting from associative attachment of detached electrons is observed. Thus, the detachment channel provides a means for background-free infrared action spectroscopy that is not otherwise measurable. The infrared spectra of both ions show clear resonances attributable to the symmetric and antisymmetric carboxylate stretching modes. The observed bands are fit to Gaussian functions to yield a band center, which is estimated to be accurate to ±5 cm$^{-1}$ despite the relatively large inherent width of the bands. The peak positions of the symmetric and antisymmetric carboxylate stretches for acetate are 1305 and 1590 cm$^{-1}$, respectively (Figure 3). In addition, the OCO bending mode is also clearly discernable at 835 cm$^{-1}$. Electron detachment from the propionate anion yields carboxylate stretching modes at 1305 and 1600 cm$^{-1}$ and the OCO bending mode at 815 cm$^{-1}$ (Figure 4).

The spectra are further analyzed using DFT and *ab initio* methods with various basis sets, requiring a confident determination of the minimum energy configurations. The geometry optimizations of the aliphatic carboxylates proceed to multiple stationary points on the potential energy surfaces, which are connected by internal rotations about the C-C bonds of acetate and propionate. The DFT optimization and frequency calculations of the acetate anion reveal two robust minima, each of C$_s$ symmetry. The energy difference between these rotamers is essentially negligible as calculated by DFT and CCSD(T) methods using the aug-cc-pVDZ basis, and the barrier to internal rotation is also calculated to be extremely small, on the order of 0.1 kJ/mol (see Supporting Information). Such differences are negligible at the temperature of the experiment (293 K) and moreover, the predicted IR spectra are very similar for the two rotamers.

The propionate B3LYP/aug-cc-pVDZ optimizations reveal four stationary points on the potential energy surface of C$_s$ symmetry, with only one robust minimum with real frequencies, shown in Figure 4. There is some disagreement between basis sets among the DFT calculations, however, as the 6-311++G** basis set produces an imaginary frequency at this geometry and optimizes to a staggered structure. CCSD(T) calculations using the 6-311++G** and aug-cc-pVDZ basis sets show the C$_s$ symmetry structure to be slightly lower in energy, but the difference between conformations is small and the effect on the predicted IRMPD spectrum is negligible. The internal rotation about the C-C bonds in the propionate anion shows a larger, but still relatively small, barrier of approximately 3 kJ/mol for rotation of the carboxylate group. The low calculated energy barrier to internal rotation of the aliphatic carboxylate anions is in sharp contrast to that of the benzoate anion, with a barrier of approximately 17 kJ/mol at the B3LYP/aug-cc-pVDZ level. Further information on the anion rotamers is available in the Supporting Information.

The vibrational spectra in Figures 3 and 4 show good overall agreement with scaled harmonic B3LYP/aug-cc-pVDZ frequency calculations. The frequencies of the symmetric and antisymmetric carboxylate stretches for benzoate, acetate and propionate are listed in Table 1. The energy difference between symmetric and antisymmetric carboxylate stretches for acetate is clearly smaller than that observed for the benzoate anion, and certainly within the error of the experiment.[15] The splitting for propionate is also smaller than that for benzoate, but slightly less so. An increase to higher values of the $COO^-$ antisymmetric stretch is commonly observed upon addition of an electron withdrawing group in aromatic[15,30] and aliphatic[16] carboxylates. Thus, the lower value for the antisymmetric stretch for the aliphatic carboxylates as compared to benzoate is interpretable as due to the relative electron donating tendency of the alkyl groups.

The large width of the spectral peaks observed for acetate and propionate is striking, but it is clear by comparison to the benzoate detachment spectrum in Figure 1 that this is not due to the detection technique. The increased width of the aliphatic carboxylates as compared to benzoate is most likely due to a greater degree of conformational fluxionality, as it generally leads to larger anharmonicity of vibrational modes.[3,4] As noted above, the calculated barriers to internal rotation are significantly different between the aliphatic and aromatic carboxylates. The large width induces some uncertainty in the assigned peak position, as it is possible that the high-energy band onset is a more appropriate indicator of the fundamental transition than the peak maximum. While demonstrating good overall agreement, the B3LYP/aug-cc-pVDZ calculations overestimate the values for the observed symmetric $COO^-$ stretch and also very slightly for the antisymmetric $COO^-$ stretch. This may be indicative of enhanced spectral activity to the red side of the IRMPD absorption bands, which is not unusual,[3] but the bands do not show enough of a significant asymmetry in shape to warrant not assigning them on the basis of peak maxima.

Gas-phase values obtained here are also compared to literature values obtained from condensed-phase experiments as shown in Table 1. It is clear that the condensed-phase experiments show significantly different values for these vibrations and the calculated values correlate well with the gas-phase experiments, which also warrants further computational investigation of the solution spectra. As shown in Table 1, the B3LYP/aug-cc-pVDZ harmonic vibrational frequency calculations performed using the Onsager dipole-dipole model of aqueous solution reproduce these shifts qualitatively if not quantitatively. The red-shift of the antisymmetric stretch in aqueous solution is accounted for well by the model, but the blue-shift in the symmetric stretch is not fully reproduced.  This trend is consistent for all the three carboxylate anions studied, suggesting improvement is possible by use of a more sophisticated reaction field model, but demonstrating a general qualitative spectroscopic agreement.  A reduced splitting between the carboxylate stretches is to be expected from solvent interaction and/or counter-ion binding.

**C. Energetics of Decarboxylation and Electron Detachment**

In addition to facilitating spectral interpretation, the calculations also provide a foundation for discussion of the relative energies of dissociation and detachment for the three carboxylate anions studied.  For the benzoate anion, the B3LYP/aug-cc-pVDZ calculation accurately reproduce the observed spectrum, so this method was chosen for generating optimum structures for higher order *ab initio* energy calculations.  The conformational analysis and spectral interpretation performed for acetate and propionate allows for a consistent model framework to treat these three anions, as the DFT calculations provide good agreement with the data and with coupled-cluster calculations.  Thus, the optimized B3LYP/aug-cc-pVDZ structure (and zero-point energy correction) from the eclipsed acetate conformer is used for CCSD(T) energy calculations and as starting point for the optimizations and frequency calculations of the neutral $CH_3COO$ radical. The $C_s$ symmetry structure is the starting point for CCSD(T) energy calculations on propionate and for optimizations of the $C_2H_5COO$ radical.

The zero-point energy-corrected electron affinities and decarboxylation energies are shown in Table 2 (further results using different basis sets are given in the Supporting

Information). The calculated (vertical and adiabatic) electron detachment energies are larger than the decarboxylation energies for all three ions. The electron affinities of the $C_2H_5COO$ and $CH_3COO$ radicals are roughly equivalent and both are smaller than the electron affinity of $C_6H_5COO$ by a few tenths of an eV. These calculated values compare favorably to experimentally determined adiabatic detachment energy values for benzoate[20], acetate[21-24] and propionate[23], as well as to previous computational results for acetate[24] and benzoate.[20]

All anions are stable with respect to $CO_2$ loss and the endothermicity of this process is similar for all three anions, but the energy of this process is somewhat greater for propionate than for acetate, and smallest for benzoate. As shown in Table 3, this can be related to the relative stabilities of the anionic fragments, since the electron affinity (EA) of $C_6H_5$ is about 1 eV, the EA of $CH_3$ is barely positive, and that of $C_2H_5$ is negative. Comparison of the electron affinities in Tables 2 and 3 shows that the carboxylate group plays a large role in stabilization of the negative charge, especially for the aliphatic carboxylate anions.

The calculated values in Table 3 compare favorably to previously reported calculated and experimental electron affinities of 1.096 eV for the $C_6H_5$ radical,[31] approximately 0.05 eV for the methyl radical[32-34] and –0.26 eV for the ethyl radical.[35] The small or even negative electron affinities of the methyl and ethyl radicals suggest that decarboxylation of the acetate and propionate parent ions would produce short-lived anions on the timescale of our experiments. Thus, while the failure to observe dissociation for acetate and propionate may be a manifestation of the slow intramolecular dynamics in systems with small vibrational state densities, it is also possible that electron detachment proceeds through an intermediate decarboxylation step, followed by prompt autodetachment of the resulting hydrocarbon anion. This mechanism would be energetically favored, however, the clear observation of electron detachment from the benzoate anion – despite the significant thermodynamic stability of $C_6H_5^-$ – suggests that the detachment process can proceed directly from the parent ion.

**IV. Conclusions**

The smallest aliphatic carboxylates acetate and propionate have been characterized in the gas phase by a novel infrared action spectroscopy method, in which IR absorption is monitored by detachment of an electron followed by capture of the electron by a scavenger, which is detected in the mass spectrum. DFT calculations confirm the benchmark gas-phase frequency values for the splitting of the symmetric and antisymmetric carboxylate stretch modes. The splitting for the aliphatic carboxylates is smaller than that for benzoate, presumably due to the greater electron withdrawing effect of the phenyl ring as compared to the alkyl groups. This effect is consistent with that observed for substituted aromatic and aliphatic carboxylates, where addition of electron-withdrawing groups increases the observed splitting of the bands. As for benzoate, the condensed-phase spectra are significantly shifted relative to the gas-phase spectra.

The method used for spectroscopic detection was validated by application to the previously characterized benzoate anion. The observation of both dissociation and detachment from benzoate suggests opportunities for investigating the competition between these processes often observed in metastable anionic species. The rates of these unimolecular reactions are related to the energetic and kinetic differences between the channels. The relative success of RRKM/QET statistical formulations suggests the possibility of modeling the competition between the dissociation and detachment within this framework.

Thermodynamic energy differences and vibrational frequencies were investigated using DFT and coupled-cluster methods. The electron detachment and decarboxylation energies are shown to be similar for the benzoate, acetate, and propionate ions, so it is suggested that the failure to observe the $CH_3^-$ or $C_2H_5^-$ dissociation products may be due to the thermodynamic instability of the fragment anions or possibly be a manifestation of the difference in dissociation dynamics for large and small molecular ions.

It is anticipated that this experimental methodology can be extended to a wide variety of molecular anions, particularly enabling infrared action spectroscopy of systems hitherto considered either too small or too rigid for IRMPD spectroscopy. As an example, we have recently reported the first gas-phase IR spectrum of the $C_{60}$ anion, which as a consequence of its high stability does not undergo dissociation under our experimental conditions.[36]

**Acknowledgment**

It is a pleasure to acknowledge the excellent support of Drs. Britta Redlich and Lex van der Meer as well as others of the FELIX staff. This work is part of the research program of FOM, which is financially supported by the Nederlandse Organisatie voor Wetenschappelijk Onderzoek (NWO).

**Supporting Information Available**

Mass spectra for IR induced electron detachment of benzoate and electron capture by $SF_6$ and $CH_3I$; extended versions of Tables 2 and 3; calculated structures for acetate and propionate; energy differences and torsional barriers between acetate and propionate rotamers. This information is available free of charge via the Internet at http://pubs.acs.org.

|  | $\nu_a$ | $\nu_s$ | $\Delta\nu_{a\text{-}s}$ |
|---|---|---|---|
| **Benzoate** | | | |
| Na salt exp. (Ref. 28) | 1553 | 1410 | 143 |
| solution (aq.) exp.(Ref. 28) | 1545 | 1390 | 155 |
| *solution (aq.) B3LYP/aug-cc-pVDZ* | *1552* | *1313* | *239* |
| gas phase exp. (Ref. 13) | 1626 | 1311 | 315 |
| *B3LYP/aug-cc-pVDZ* | *1627* | *1311* | *316* |
| **Acetate** | | | |
| Na salt exp. (Ref. 14) | 1583 | 1421 | 162 |
| solution (aq.) exp.(Ref. 15) | 1551 | 1416 | 135 |
| solution (aq.) exp.(Ref. 16) | 1552 | 1415 | 137 |
| *solution (aq.) B3LYP/aug-cc-pVDZ* | *1554* | *1328* | *226* |
| gas phase exp. (this work) | 1590(10) | 1305(10) | 285(14) |
| *B3LYP/aug-cc-pVDZ* | *1603* | *1321* | *282* |
| **Propionate** | | | |
| Na salt exp. (Ref. 14) | 1565 | 1429 | 136 |
| solution (aq.) exp.(Ref. 17) | 1545 | 1413 | 132 |
| *solution (aq.) B3LYP/aug-cc-pVDZ* | *1557* | *1327* | *231* |
| gas phase exp. (this work) | 1600(10) | 1305(10) | 295(14) |
| *B3LYP/aug-cc-pVDZ* | *1604* | *1317* | *287* |

**Table 1.** Observed values in cm$^{-1}$ for the symmetric and antisymmetric carboxylate stretching frequencies in condensed and gas phases as compared to DFT calculations for the benzoate, acetate and propionate anions. All calculated harmonic frequencies are scaled by a factor of 0.98. Calculated frequencies in solution phase use the Onsager dipole model for SCF convergence. $\nu_a$ and $\nu_s$ denote the COO$^-$ antisymmetric and symmetric stretch modes.

|  | Benzoate | | | Acetate | | | Propionate | | |
|---|---|---|---|---|---|---|---|---|---|
|  | Diss. | AEA | VDE | Diss. | AEA | VDE | Diss. | AEA | VDE |
| *B3LYP/* | | | | | | | | | |
| aug-cc-pVDZ | 2.39 | 3.42 | 3.95 | 2.54 | 3.13 | 3.64 | 2.75 | 3.16 | 3.59 |
| aug-cc-pVTZ | 2.41 | 3.43 | 3.96 | 2.41 | 3.14 | 3.63 | 2.61 | 3.17 | 3.58 |
| 6-31+G** |  | 3.44[a] |  |  |  |  |  |  |  |
| *CCSD(T)/* | | | | | | | | | |
| 6-311+G* | 2.31 | 3.30 | 3.78 | 2.63 | 2.93 | 3.35 | 2.87 | 2.98 | 3.47 |
| 6-311++G** | 2.27 | 3.33 | 3.80 | 2.51 | 2.98 | 3.39 | 2.74 | 3.03 | - |
| aug-cc-pVDZ | - | - | - | 2.55 | 3.12 | 3.54 | 2.84 | 3.16 | - |
| aug-cc-pVTZ |  |  |  |  | 3.33[b] |  |  |  |  |

[a] Ref. 20
[b] Ref. 24

**Table 2.** Calculated decarboxylation and electron detachment energies for acetate, propionate, and benzoate anions. The energy of $CO_2$ loss (Diss.) is compared to the adiabatic electron affinity (AEA) of the corresponding radical and the vertical electron detachment energy (VDE) of the anion. All values are given in eV, and are zero-point energy corrected. Single-point CCSD(T) energy calculations are performed at B3LYP/aug-cc-pVDZ optimized structures and are zero-point energy corrected using the frequencies at this level of theory.

|  | Adiabatic Electron Affinities | | |
| --- | --- | --- | --- |
|  | $C_6H_5$ | $CH_3$ | $C_2H_5$ |
| *B3LYP/* | | | |
| aug-cc-pVDZ | 1.09 | -0.004 | -0.19 |
| aug-cc-pVTZ | 0.94 | -0.004 | -0.20 |
| | | | |
| *CCSD(T)/* | | | |
| 6-311++G** | 1.03 | 0.010 | -0.48 |
| aug-cc-pVDZ | - | 0.002 | -0.30 |
|  | 1.096[a] | 0.05[b] | -0.26[d] |
|  |  | 0.044[c] |  |

[a]Experimental value from photo-electron spectrum (Ref. 31)
[b]Ref. 33
[c]Ref. 34
[d]Value derived from thermodynamic cycle (Ref. 35)

**Table 3.** Calculated electron affinities of $CH_3$, $C_2H_5$, and $C_6H_5$. All values are given in eV, and are zero-point energy corrected at the level of theory stated, except for CCSD(T) calculations, which are performed at B3LYP/aug-cc-pVDZ optimized structures and use the zero-point energy derived from frequencies also at this level of theory.

**List of Figure Captions**

**Figure 1:** IR multiple photon excitation spectrum of the benzoate anion detected simultaneously in the dissociation channel (*m/z* 77) and the detachment/attachment channel (to $SF_6$, *m/z* 146). Also shown is the depletion of the benzoate parent ion, (*m/z* 121).

**Figure 2:** Comparison of detachment spectra of the carboxylate infrared bands of the benzoate anion obtained using different electron scavengers, $SF_6$ and $CH_3I$. Associative electron attachment to $SF_6$ produces $SF_6^-$ ions, and dissociative electron attachment to methyl iodide results in $I^-$ formation. Example mass spectra for the two scavenger experiments are available in the Supporting Information.

**Figure 3.** IR multiple photon detachment/attachment spectra of the acetate anion. Also shown is the depletion of the parent ion, (*m/z* 59). Under our experimental conditions, no dissociation was observed for these species. The symmetric and antisymmetric carboxylate stretching modes are clearly observed and correspond well to computed spectra at the B3LYP/aug-cc-pVDZ level. The lowest-energy calculated structure at this level of theory is also shown.

**Figure 4.** IR multiple photon detachment/attachment spectra of the propionate anion. Also shown is the depletion of the parent ion, (*m/z* 73). As for acetate, the symmetric and antisymmetric carboxylate stretching modes are clearly observed and correspond well to computed spectra at the B3LYP/aug-cc-pVDZ level. The lowest-energy calculated structure at this level of theory is also shown.

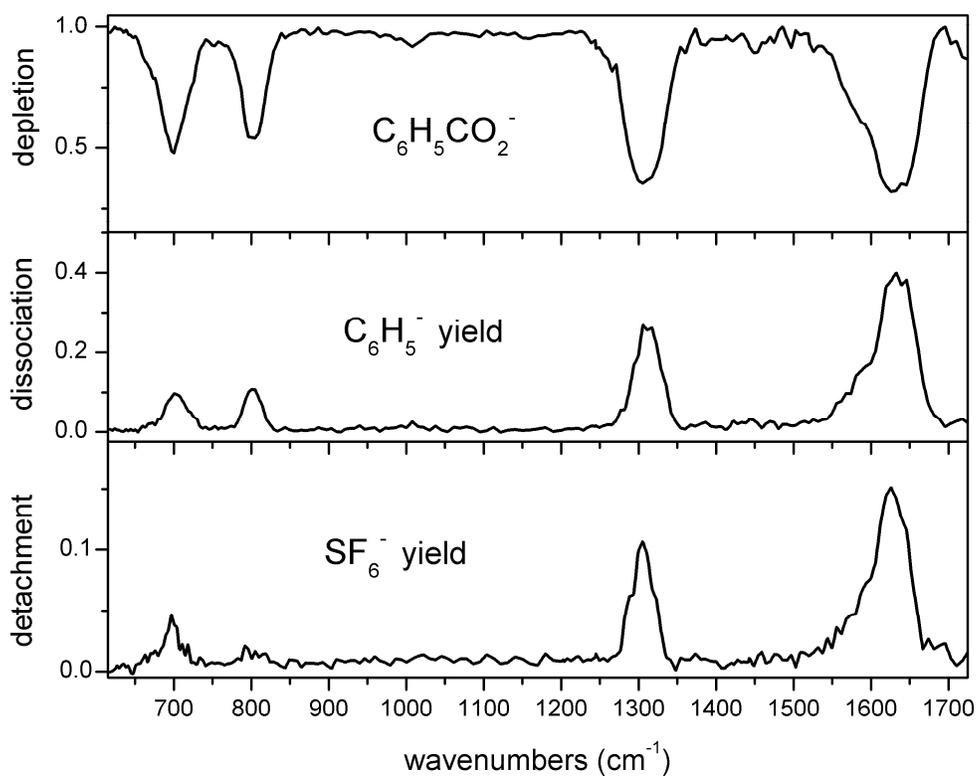

**Figure 1**

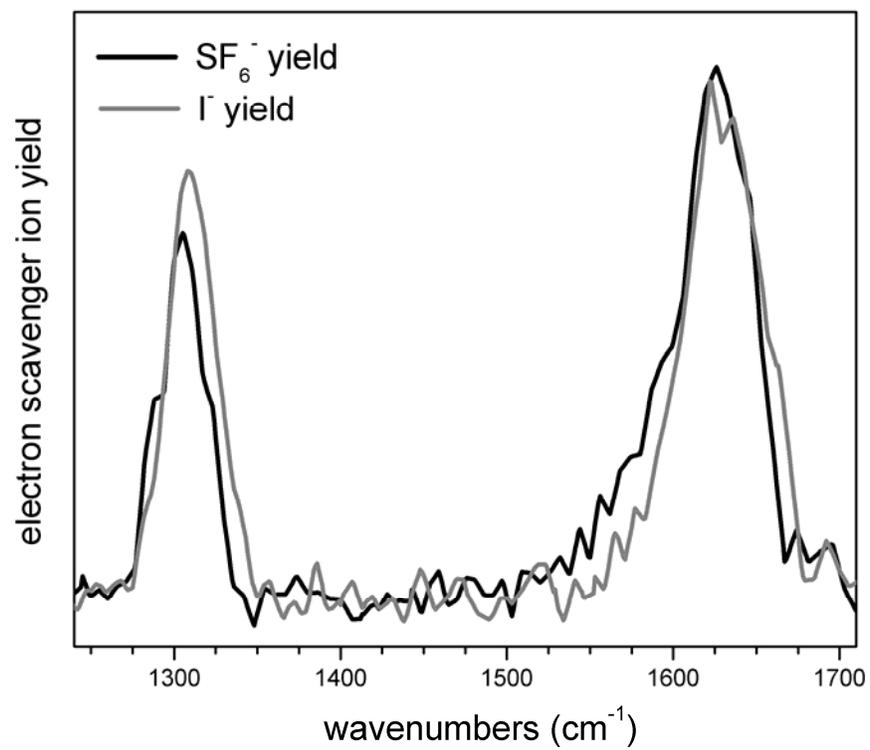

**Figure 2**

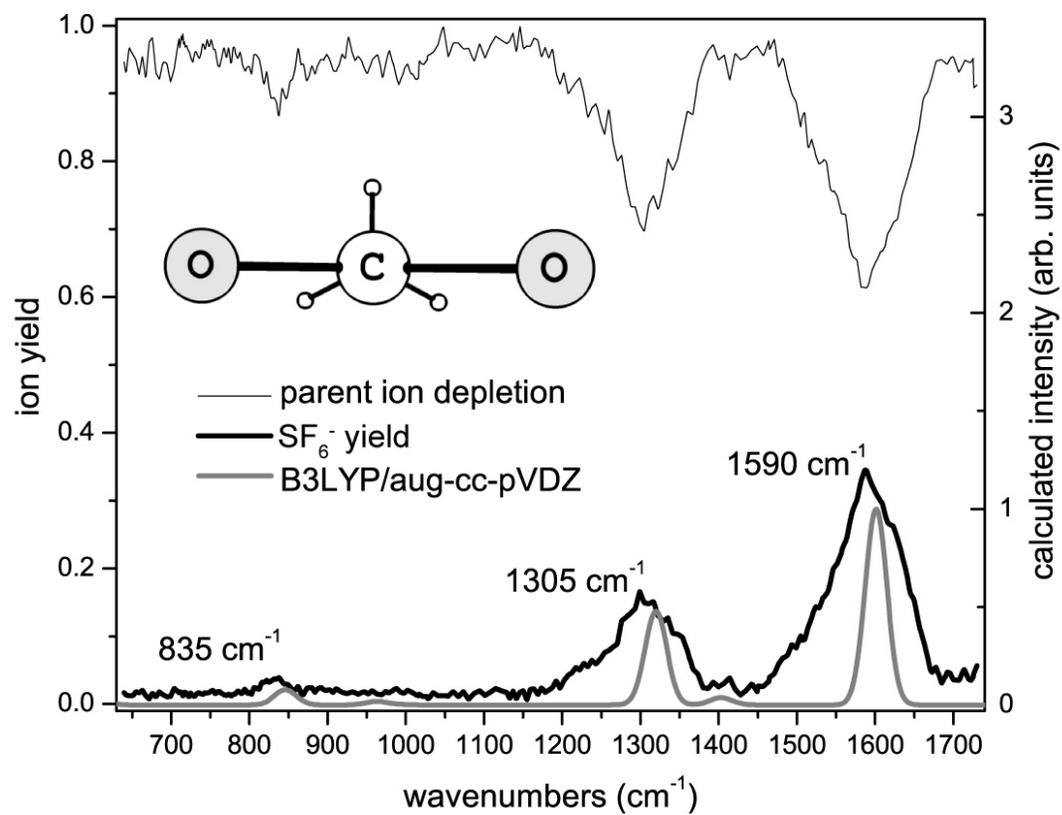

**Figure 3**

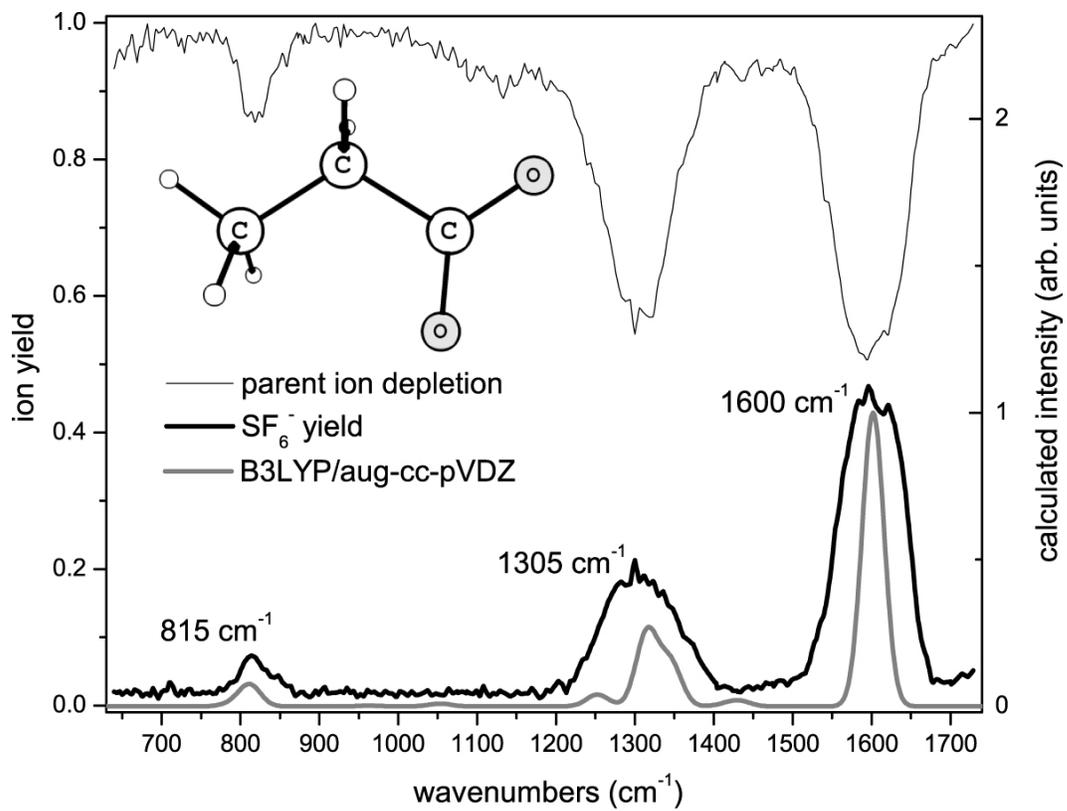

**Figure 4**

**Table of contents entry:**

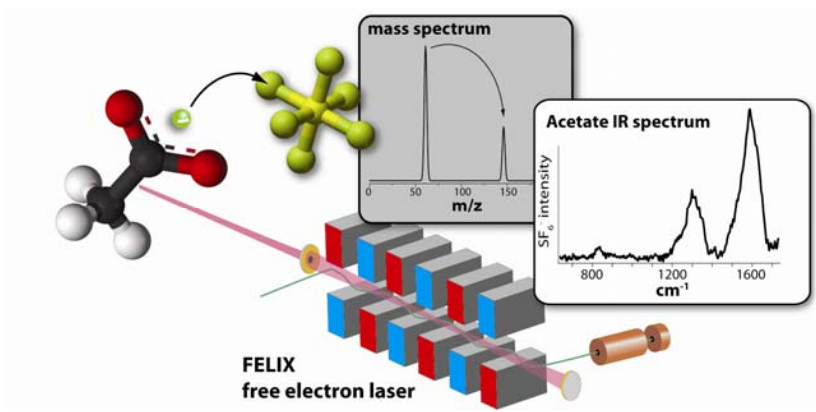

IR induced electron detachment and subsequent capture by a scavenger is used as 'action spectroscopy' method for anions.